\documentclass[journal=apchd5,manuscript=article]{achemso}
\usepackage[utf8]{inputenc}

\usepackage{graphicx}
\usepackage{amsfonts}
\usepackage{amssymb}
\usepackage{amsmath}
\usepackage{doi}
\usepackage{setspace}
\title{Inverse-designed photonics for semiconductor foundries}

\author{Alexander Y. Piggott}
\altaffiliation{These authors contributed equally to the work.}
\affiliation{Ginzton Laboratory, Stanford University, Stanford, CA, USA}
\author{Eric Y. Ma}
\affiliation{Ginzton Laboratory, Stanford University, Stanford, CA, USA}
\alsoaffiliation{Stanford Institute for Materials and Energy Sciences, SLAC National Accelerator Laboratory, Menlo Park, CA, USA}
\altaffiliation{These authors contributed equally to the work.}
\author{Logan Su}
\affiliation{Ginzton Laboratory, Stanford University, Stanford, CA, USA}
\altaffiliation{These authors contributed equally to the work.}
\author{Geun Ho Ahn}
\author{Neil V. Sapra}
\author{Dries Vercruysse}
\affiliation{Ginzton Laboratory, Stanford University, Stanford, CA, USA}
\author{Andrew M. Netherton}
\author{Akhilesh S.P. Khope}
\author{John E. Bowers}
\affiliation{University of California Santa Barbara, Santa Barbara, CA, USA}
\author{Jelena Vu\v{c}kovi\'c}
\affiliation{Ginzton Laboratory, Stanford University, Stanford, CA, USA}
\email{jela@stanford.edu}

\begin{document}


\begin{tocentry}

	\centering
	\includegraphics[width=6.5cm]{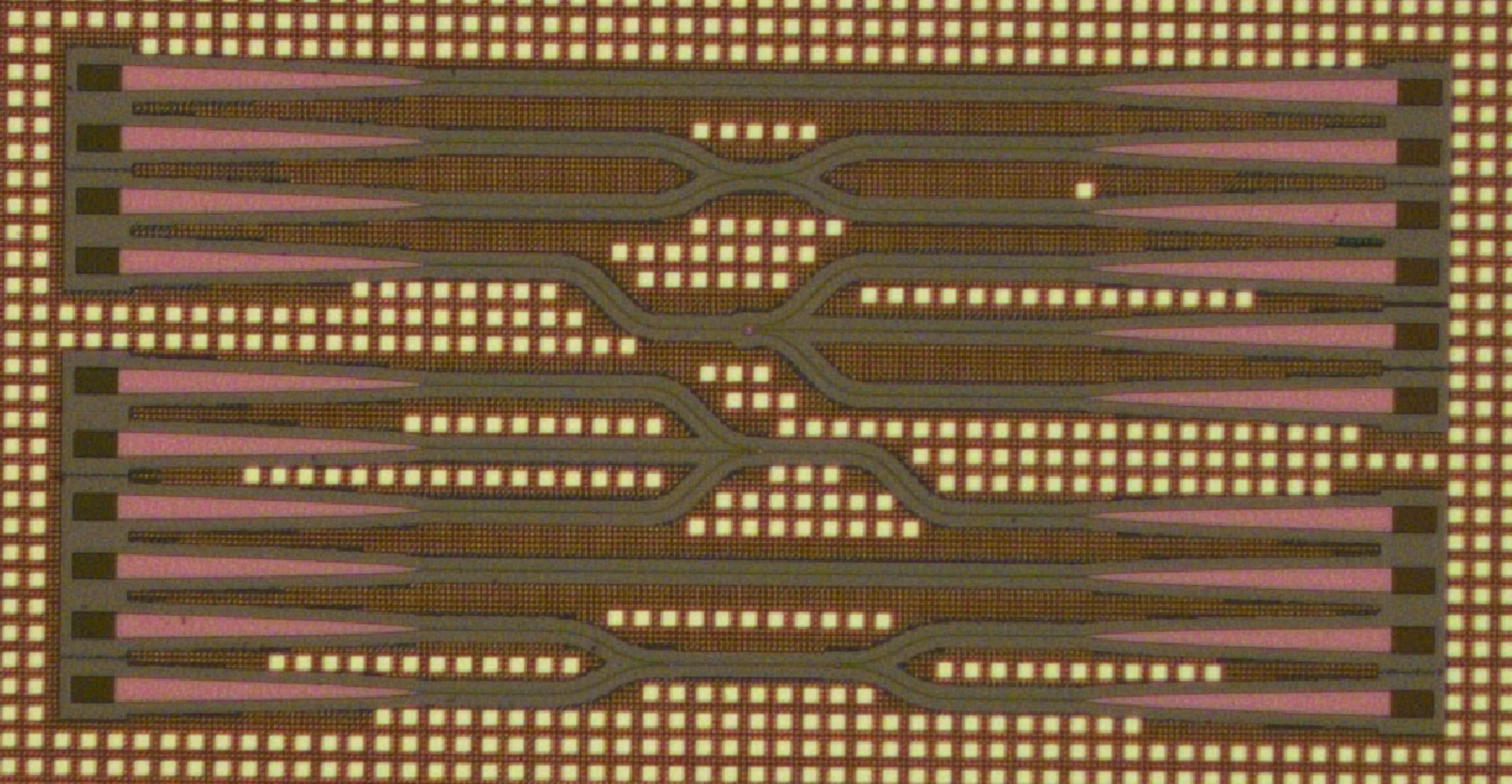}

\end{tocentry}

\begin{abstract}
Silicon photonics is becoming a leading technology in photonics, displacing traditional fiber optic transceivers and enabling new applications. Further improving the density and performance of silicon photonics, however, has been challenging, due to the large size and limited performance of traditional semi-analytically designed components. Automated optimization of photonic devices using inverse design is a promising path forward but has until now faced difficulties in producing designs that can be fabricated reliably at scale. Here we experimentally demonstrate four inverse-designed devices - a spatial mode multiplexer, wavelength demultiplexer, 50-50 directional coupler, and 3-way power splitter - made successfully in a commercial silicon photonics foundry. These devices are efficient, robust to fabrication variability, and compact, with footprints only a few micrometers across. They pave the way forward for the widespread practical use of inverse design.

\end{abstract}

Silicon photonics is becoming a leading technology in photonics \cite{siph_yole2019} by displacing traditional photonics and enabling new applications in a wide variety of product areas. For example, silicon photonic transceivers are quickly becoming the de-facto standard for fiber optics links, ranging from long-haul telecommunications to intra-data-center links  \cite{arahim_pieee2018}. New applications such as LiDAR (Light Detection And Ranging) \cite{jsun_nat2013, samiller_cleo2018} and optical machine learning \cite{yshen_np2017,komljenovic2016heterogeneous} are actively being developed. The key to the success of silicon photonics is that it leverages standard CMOS (Complementary Metal Oxide Semiconductor) fabrication processes, allowing high-performance optical systems to be produced in large volumes at very low cost \cite{komljenovic2016heterogeneous,gtreed_2008}.

Progress in silicon photonics, however, has long been hampered by the small library of semi-analytically designed devices in common use. These traditional designs are rather large, ranging from tens to hundreds of microns in size for even basic functions, and often leave much to be desired in terms of performance and robustness. A promising solution is inverse design, whereby photonic devices are designed by optimization algorithms with little-to-no human input \cite{amutapcica_eo2009, jjensen_lpr2011, lalau-keraly_oe2013, jlu_oe2013, aniederberger_oe2014, aypiggott_np2015, lffrellsen_oe2016, sell2017large, amichaels_oe2018, molesky2018inverse}. Inverse design has successfully produced designs that have improved optical performance, improved robustness to errors in fabrication and variation of operational conditions, or use orders of magnitude less area, when compared to previous designs. Such compact devices are especially useful for newer applications of silicon photonics that require high photonic component densities, such as phased arrays for LiDAR systems and dense arrays of Mach-Zehnder interferometers for machine learning. Unfortunately, the devices generated by inverse design often have small features that are difficult to fabricate reliably using photolithography, the mainstay of commercial semiconductor manufacturing. Indeed, the vast majority of previous experimental demonstrations of inverse-designed photonics have used either electron-beam lithography or focused ion beam machining, which have considerably higher resolution but cannot be used to produce devices at scale \cite{jjensen_lpr2011, aypiggott_np2015, lffrellsen_oe2016, aypiggott_sr2017, lsu_acsphoton2018, cdory_nc2019}. There are a few demonstrations using photolithography, but they unnecessarily restrict the design space and cannot handle arbitrary topologies \cite{hoffman2019improved, mak2016binary}. To the best of our knowledge, the only previous attempt at fabricating inverse-designed devices at a foundry had poor agreement between simulated and experimental performance due to significant differences between the designed and fabricated structures \cite{mak2016binary}.

\begin{figure}
    \centering
    \includegraphics[width=\textwidth]{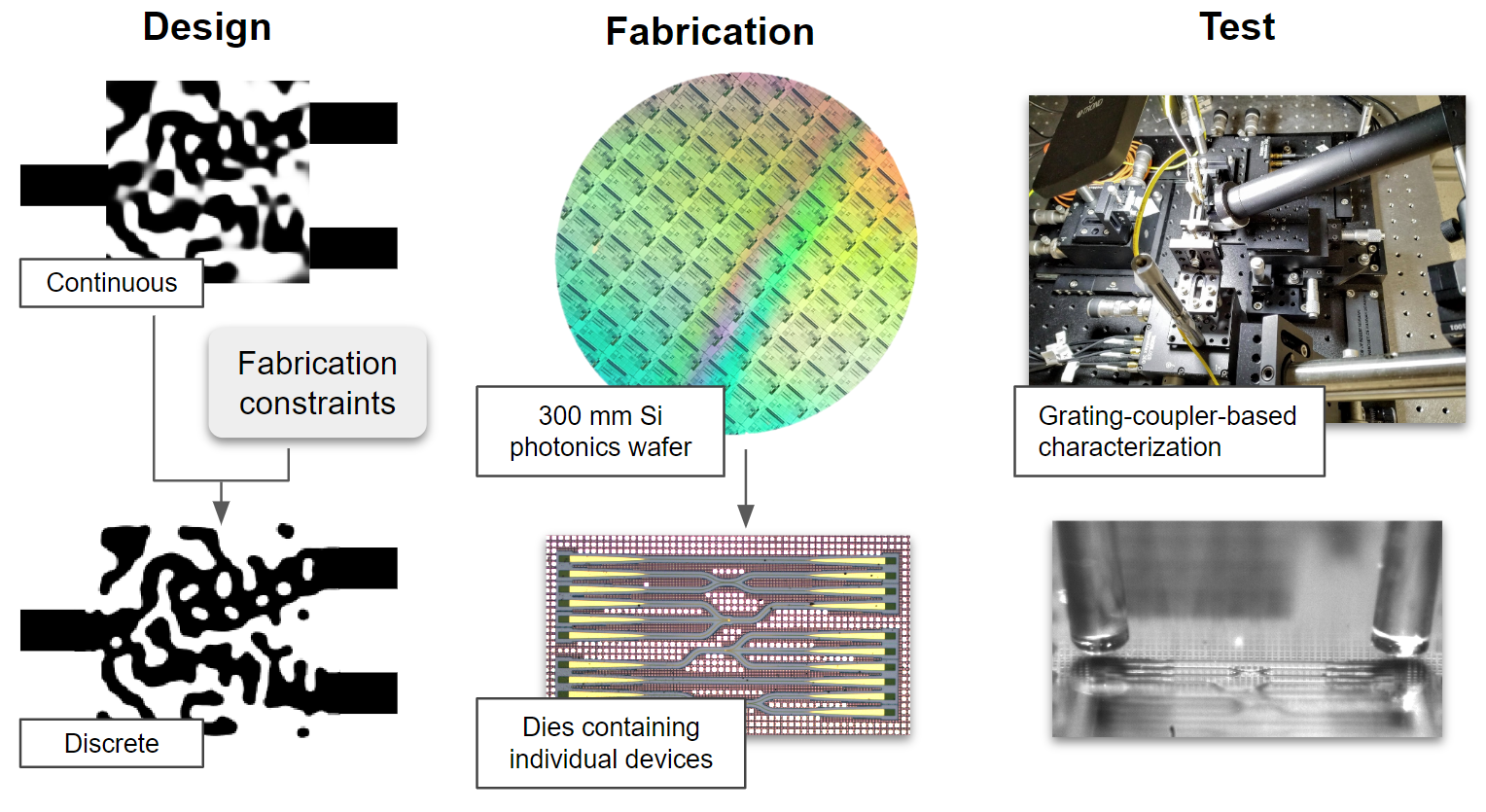}
    \caption{Workflow of photonics inverse design for commercial silicon photonics foundries. Fabrication constraints are applied in the discrete optimizations. The final device design is then combined with components from the foundry's PDK (Process Design Kit), e.g. waveguides and grating couplers, to complete the final mask pattern. AIM Photonics 300 mm multi-project Si wafers are then fabricated via water-immersion deep UV photolithography at the Albany NanoTech fabrication facility. The wafer is diced and the devices tested in a vertical transmission measurement setup. (*Wafer image by Frank Tolic. Other photos taken by authors.)}
    \label{fig:foundry_overview}
\end{figure}

In this work, we report the first successful demonstration of inverse-designed photonics in a commercial silicon photonics process. The designs were fabricated as part of the AIM Photonics 300 mm wafer multi-project wafer (MPW) foundry offering \cite{aimphotonics}. We demonstrate four different devices fabricated using a single fully-etched layer of $220~\mathrm{nm}$ thick silicon, surrounded on all sides by silicon dioxide cladding. These devices are compact, with footprints of only several micrometers across, and have comparable performance and reproducibility to previous inverse-designed devices fabricated using electron-beam lithography (Fig. \ref{fig:foundry_overview}).

\section*{Results}
\subsection*{Design}
The goal of inverse design is to automate the design process. First, a human designer broadly specifies the desired performance and other characteristics of the device, such as the desired transmission through an output port with a given design area \cite{aypiggott_sr2017, lsu_acsphoton2018, aypiggott_phdthesis}. An optimization algorithm is then used to search the space of available designs using gradient-based optimization, which can efficiently optimize over tens or even hundreds of thousands of design degrees of freedom. By using adjoint sensitivity analysis, the gradient can be efficiently computed using one additional electromagnetic simulation, regardless of the number of design parameters \cite{aniederberger_oe2014, aypiggott_phdthesis, mbgiles_ftc2000, sgjohnson_adjoint2012}.


To successfully fabricate devices at a foundry, the designs should be robustly resolved using photolithography. In principle, it would be possible to directly incorporate a lithography model into the optimization algorithm, but this requires detailed knowledge of the lithography parameters used by the foundry. In lieu of directly incorporating a lithography model, \cite{aypiggott_sr2017} proposed using two constraints as heuristics: a minimum gap and a minimum radius of curvature. More specifically, a minimum radius of curvature constraint is applied to all material interfaces, preventing the formation of any sharp cusps and corners. A minimum gap constraint prevents the formation of narrow gaps and bridges. To ensure robustness to fabrication errors, all devices were designed to operate over as broad of a range of wavelengths as possible, which has previously been shown to be an effective heuristic for fabrication robustness. As will be demonstrated through the experimental measurements, these two fabrication constraints, along with broadband optimization, are sufficient for creating devices that can be reliably fabricated at commercial foundries.

\subsection*{Spatial mode multiplexer}
\begin{figure}
    \centering
    \includegraphics[width=\textwidth]{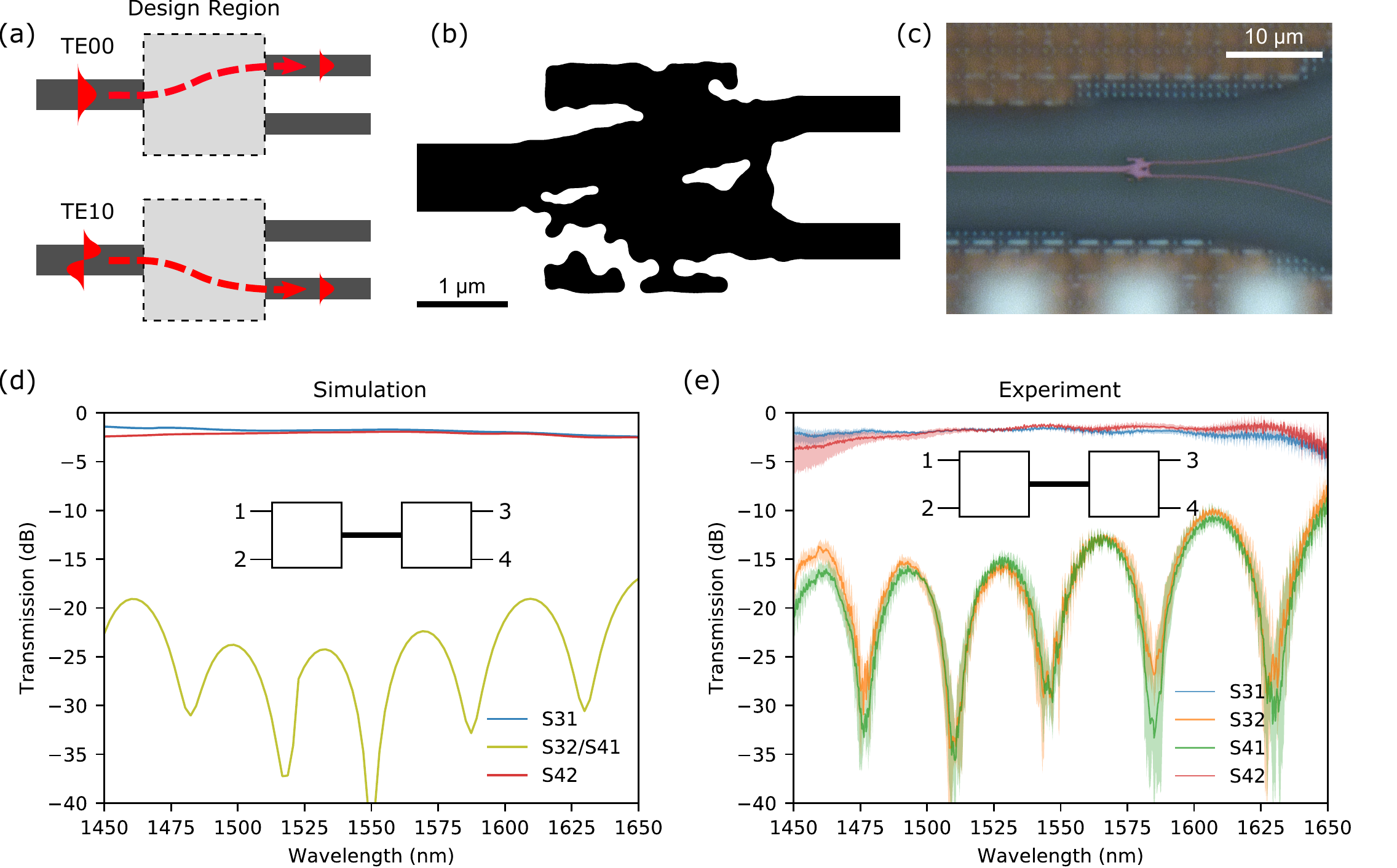}
    \caption{A spatial mode multiplexer. (a) The spatial mode multiplexer maps the $\mathrm{TE}_{00}$ and $\mathrm{TE}_{10}$ modes of the $750~\mathrm{nm}$ wide input waveguide to the $\mathrm{TE}_{00}$ mode of the two $400~\mathrm{nm}$ wide output waveguides. (b) The final design, with regions of silicon indicated by black, and silicon dioxide indicated by white. (c) An optical microscopy image of the final fabricated device. (d) Simulated and (e) experimentally measured S-parameters for the back-to-back test structure that allows the design to be measured using only standard single-mode optical waveguides. The shaded areas in (e) indicate the minimum and maximum values across three different measured devices from three dies, and the solid lines the average. }
    \label{fig:foundry_wgSDM}
\end{figure}
First we consider a compact spatial mode multiplexer, which separates the fundamental $\mathrm{TE}_{00}$ and second-order $\mathrm{TE}_{10}$ modes of a $750~\mathrm{nm}$ wide multi-mode input waveguide, and routes them to separate $400~\mathrm{nm}$ wide single-mode output waveguides (Fig. \ref{fig:foundry_wgSDM}(a)). The device was designed by first allowing the permittivity in the design region to continuously vary between that of silicon and silicon dioxide, before applying thresholding and switching to boundary optimization \cite{aypiggott_sr2017}. During boundary optimization, a 70 nm minimum radius of curvature and 90 nm minimum gap constraint were applied. This resulted in a design with a complex and non-intuitive topology and a compact footprint of $3.55 \times 2.55 ~\mathrm{\mu m}^2$ (Fig. \ref{fig:foundry_wgSDM}(b, c)).

To test the spatial mode multiplexer, two multiplexers were placed back-to-back, joined by an $80~\mathrm{\mu m}$ segment of multi-mode waveguide. This allowed the device to be measured using standard single-mode optical fibers and grating couplers. Fig. \ref{fig:foundry_wgSDM}(d) and (e) show that the simulated and measured S-parameters agree well with each other. In addition, the fabricated devices are very reproducible: the S-parameters of the three instances from three dies are closely aligned. Extracting the S-parameters of a single device is straightforward (Methods): over the entire operating bandwidth of $1500 - 1600~\mathrm{nm}$, the insertion loss is $< 1.0~\mathrm{dB}$, and the crosstalk suppression is $> 15.6~\mathrm{dB}$.

\subsection*{Wavelength demultiplexer}

\begin{figure}
    \centering
    \includegraphics[width=\textwidth]{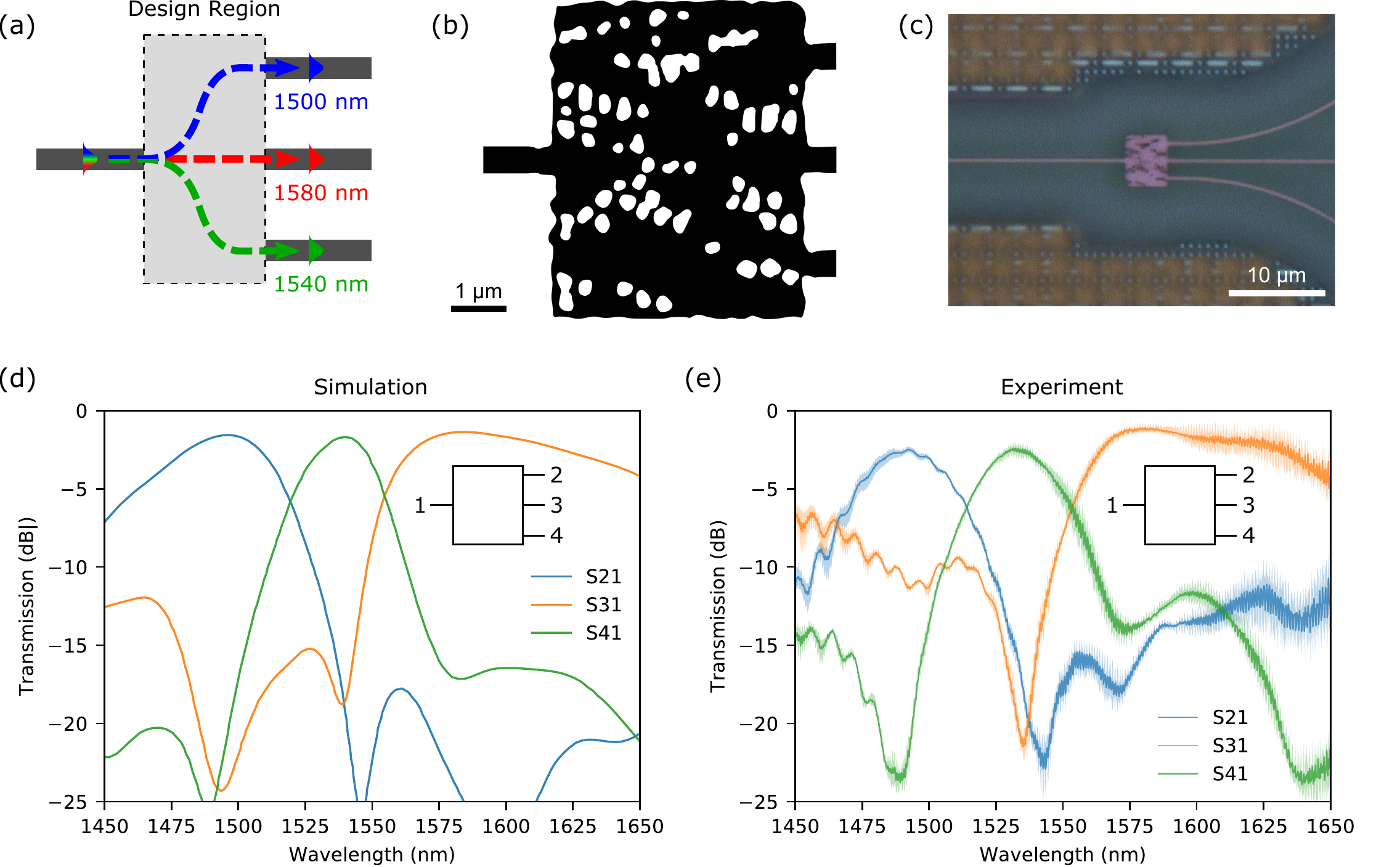}
    \caption{A 3-channel wavelength demultiplexer. (a) The wavelength demultiplexer splits $1500~\mathrm{nm}$, $1540~\mathrm{nm}$, and $1580~\mathrm{nm}$ light into three separate output waveguides. (b) The final design, with regions of silicon indicated by black, and silicon dioxide indicated by white. (c) An optical microscopy image of the final fabricated device. (d) Simulated and (e) experimentally measured S-parameters for the wavelength demultiplexer. The shaded areas in (e) indicate the minimum and maximum values across three different measured devices. The three pass-bands at $1500~\mathrm{nm}$, $1540~\mathrm{nm}$, and $1580~\mathrm{nm}$ are clearly visible in the data.} 
    \label{fig:foundry_wgWDM3}
\end{figure}
Next we consider a 3-channel wavelength demultiplexer, designed to separate $1500~\mathrm{nm}$, $1540~\mathrm{nm}$, and $1580~\mathrm{nm}$ light (Fig. \ref{fig:foundry_wgWDM3}(a)). We used ``neighbour biasing'' in the continuous stage of optimization to produce a good starting point for boundary optimization \cite{lsu_acsphoton2018}. The minimum radius of curvature was $40~\mathrm{nm}$, and the minimum gap width was $90~\mathrm{nm}$.  The design (Fig. \ref{fig:foundry_wgWDM3}(b, c)) is highly non-intuitive but compact, with a footprint of only $5.5 \times 4.5~\mathrm{\mu m}^2$.

The simulated and measured S-parameters are presented in Fig. \ref{fig:foundry_wgWDM3}(d) and (e). The measured spectra exhibit three clear passbands, showing that the device functions as intended, although the crosstalk is somewhat higher. The reproducibility is also excellent: the three fabricated instances have nearly identical transmission. The insertion loss for the 3 output channels are $3.0~\mathrm{dB}$ at $1500~\mathrm{nm}$, $3.1~\mathrm{dB}$ at $1540~\mathrm{nm}$, and $1.2~\mathrm{dB}$ at $1580~\mathrm{nm}$. The crosstalk suppression are $8.3, 12.6$ and $12.3~\mathrm{dB}$, respectively. 

\subsection*{Directional coupler}
\begin{figure}
    \centering
    \includegraphics[width=\textwidth]{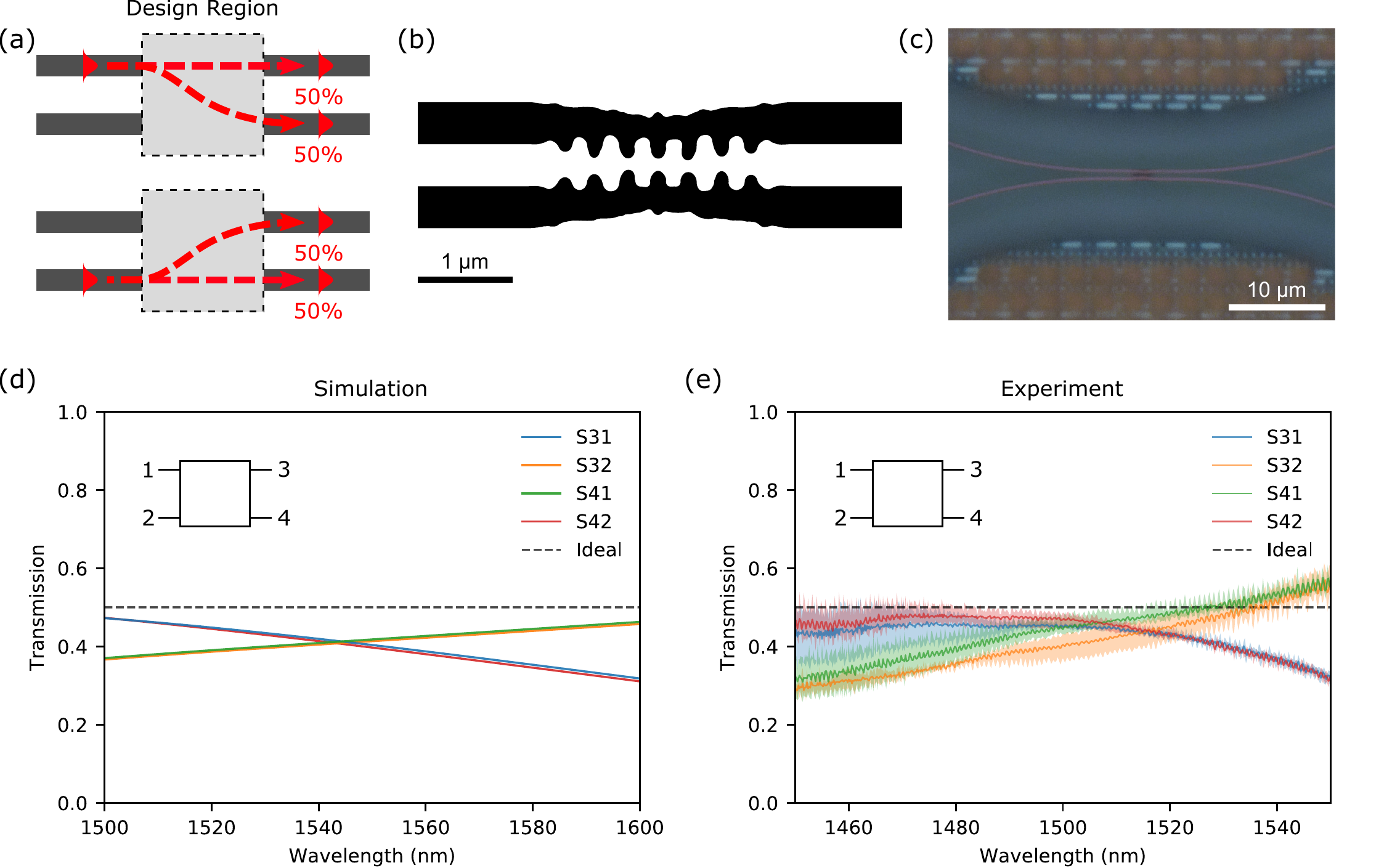}
    \caption{A 50-50 directional coupler. (a) A 50-50 directional coupler equally splits light from its two input waveguides into its two output waveguides. (b) The final design, with regions of silicon indicated by black, and silicon dioxide indicated by white. (d) Simulated and (e) experimentally measured performance of the directional coupler. The shaded areas in (e) indicate the minimum and maximum values across three different measured devices. The dashed line indicates a perfect $1/2$ splitting ratio. }
    \label{fig:foundry_wg5050dc}
\end{figure}
The third design is a 50-50 directional coupler that takes light from either of its two input waveguides and equally divides it between the two output waveguides (Fig. \ref{fig:foundry_wg5050dc}(a)). The same two-stage design process was used as for the spatial mode multiplexer, with minimum radius of curvature of $70~\mathrm{nm}$ and minimum gap of $90~\mathrm{nm}$. With a footprint of $3.0 \times 1.2~\mathrm{\mu m}^2$, the final design is significantly more compact than most designs in the literature (Fig. \ref{fig:foundry_wg5050dc}(b)). Interestingly, the structure strongly resembles a conventional grating-assisted directional coupler, despite the complete lack of human intervention throughout the design process.

The design is relatively broadband, with reasonably matched output powers over a $100~\mathrm{nm}$ bandwidth in both simulation (Fig. \ref{fig:foundry_wg5050dc}(d)) and measurement (Fig. \ref{fig:foundry_wg5050dc}(e)). Over a $45~\mathrm{nm}$ bandwidth, the fabricated couplers have an average insertion loss of $0.5~\mathrm{dB}$ and $< 10\%$ power imbalance. There is, however, a significant wavelength shift between the simulated and measured devices, likely due to fabrication errors: the design has a central wavelength of $1545~\mathrm{nm}$, whereas the measured devices operate around $1505~\mathrm{nm}$. 

\subsection*{Power splitter}

\begin{figure}
    \centering
    \includegraphics[width=\textwidth]{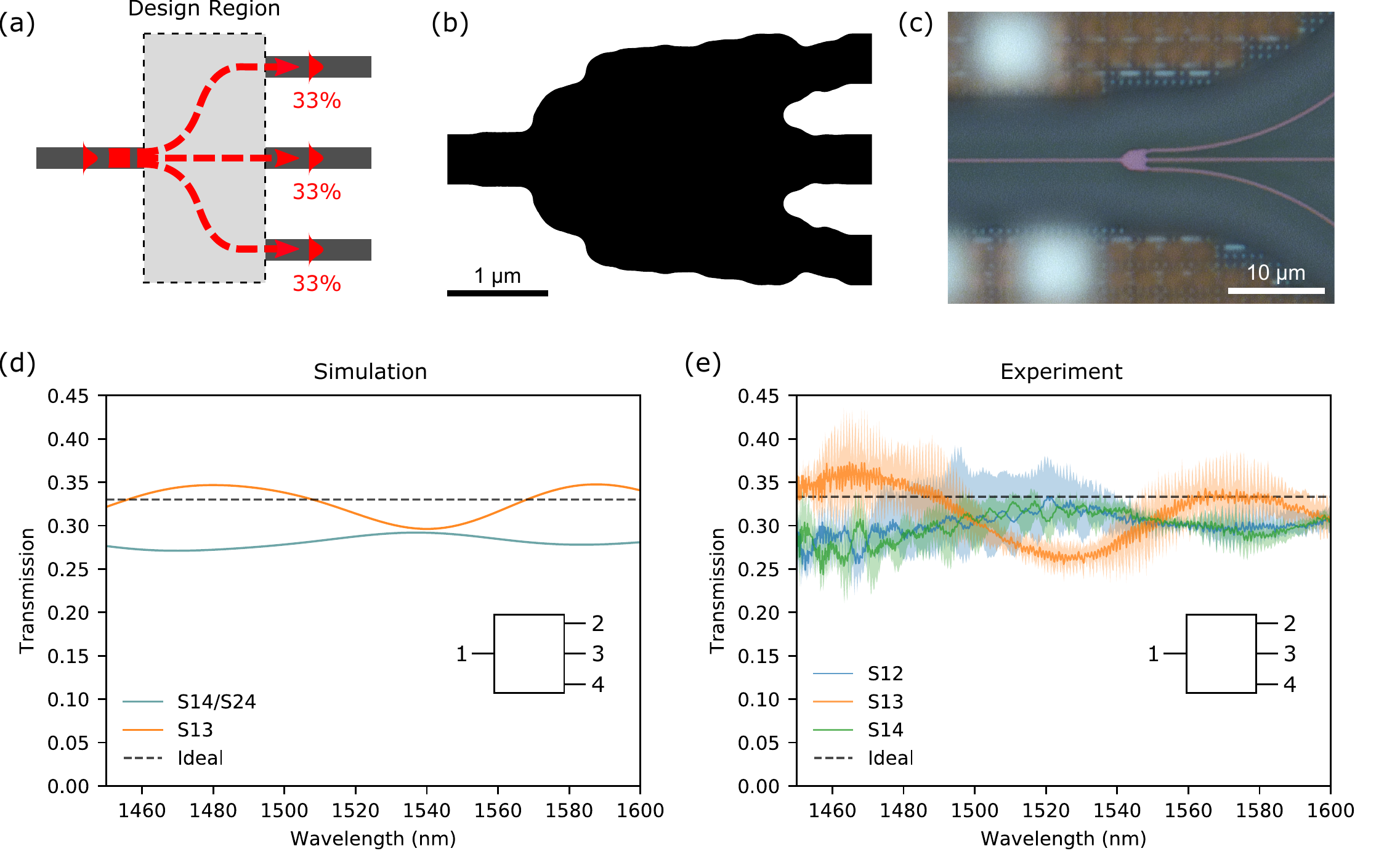}
    \caption{A 3-way power splitter. (a) The 3-way power splitter equally splits power between three output waveguides. (b) The final design, with regions of silicon indicated by black, and silicon dioxide indicated by white. (d) Simulated and (e) experimentally measured S-parameters for the 3-way power splitter. The shaded areas in (e) indicate the minimum and maximum values across three different measured devices. The dashed line indicates a perfect $1/3$ splitting ratio.}
    \label{fig:foundry_wgSplit3}
\end{figure}

The last design is a broadband three-way power splitter that equally splits the power from an input waveguide into three output waveguides (Fig. \ref{fig:foundry_wgSplit3}(a)). In contrast to the previous designs, the power splitter was designed using only boundary optimization with minimum radius of curvature of $100~\mathrm{nm}$, yielding a design (Fig. \ref{fig:foundry_wgSplit3}(b)) that resembles a compact multi-mode interferometer (MMI) coupler \cite{aypiggott_sr2017}. The final design has a footprint of $3.8 \times 2.5 ~\mathrm{\mu m}$. 

The simulated and measured S-parameters for the three-way power splitter are given in Fig. \ref{fig:foundry_wgSplit3}(d) and (e). The splitting is very broadband, operating nominally from $1450~\mathrm{nm}$ to $1600~\mathrm{nm}$. The simulated and measured S-parameters match quite closely, although there is a shift in the transmission fringes. The design was constrained to have reflection symmetry across the horizontal axis, resulting in $\mathrm{S}_{12}$ and $\mathrm{S}_{14}$ parameters that are identical in simulation and nearly identical in measurement. Over the full operating bandwidth, the splitter has an insertion loss of $0.4~\mathrm{dB}$, and a power imbalance of $4.4\%$.

\section*{Discussions}
All four designs appear to be quite robust to fabrication errors. The device-to-device variability across the three dies, each containing a single instance of all designs, was approximately equal to our measurement uncertainty of $\pm 0.6~\mathrm{dB}$, limited mostly by grating coupler variability. This consistency suggests that they are robust to typical fabrication errors, such as defocusing in photolithography or variation in layer thicknesses. Furthermore, both the wavelength demultiplexer \cite{lsu_acsphoton2018} and 3-way power splitter \cite{aypiggott_sr2017} were previously fabricated using electron-beam lithography and had comparable performance to the present devices. This implies that high-resolution electron-beam lithography is not necessary for fabricating these complex designs; industry-standard photolithography is sufficient, so long as the designs are properly constrained.

In summary, we have experimentally demonstrated a photonic inverse design process that is compatible with industry-standard photonics foundries. By incorporating fabrication constraints we have eliminated small features that cannot be resolved using photolithography. A wide breadth of devices were demonstrated, illustrating the flexibility of the method. These results show that inverse design is a suitable method for designing practical integrated photonic devices and has the potential to revolutionize the field by enabling a new generation of exceedingly compact and high performance devices.

\section*{Methods}

\subsection*{Design algorithms}
All devices were designed using our adjoint optimization based implementation of photonic inverse design \cite{aypiggott_sr2017, lsu_acsphoton2018, aypiggott_phdthesis}. Our inverse design algorithm proceeds in two stages. In the first stage, the permittivity is allowed to continuously vary between those of the available materials (e.g. silicon and silicon dioxide) at every point in the design. The design is then thresholded to produce a binary structure consisting of only two materials, which is used as the starting condition for the next stage. In the second stage, only the material boundaries are optimized by using a level set representation of the structure \cite{sosher_2003}. Under the level set representation, the boundary of the device is defined as the zero crossing of a continuous function \cite{aypiggott_sr2017}. The implicit nature of the level set representation makes it trivial to handle changes in topology, such as the merging or splitting of holes, and does not require one to perform complex additional steps such as re-meshing.

The computational cost of inverse design is dominated by the required electromagnetic simulations, which were performed using Maxwell FDFD \cite{wshin_jcp2012, wshin_oe2013}, a GPU-based implementation of the finite-difference frequency-domain (FDFD) method, with a spatial step size of $40~\mathrm{nm}$. The Maxwell FDFD simulation software is available on GitHub at \url{https://github.com/stanfordnqp/maxwell-b} under the GNU General Public License v3.0. Final broadband verification simulations were then performed using commercial Lumerical FDTD (finite-difference time-domain) software \cite{lumerical_fdtd}. All simulations and design were performed on a server with an Intel Core i7-5820K processor, 64GB of RAM, and three Nvidia Titan Z graphics cards.

The spatial mode multiplexer, 3-way power splitter, and 50-50 directional coupler were optimized over 6 equally spaced wavelengths from $1400~\mathrm{nm}$ to $1700~\mathrm{nm}$ \cite{aypiggott_sr2017}. This resulted in designs which operated well over a broad wavelength range.  Meanwhile, the wavelength demultiplexer was optimized at the three channel wavelengths of $1500~\mathrm{nm}$,  $1540~\mathrm{nm}$, and $1580~\mathrm{nm}$ \cite{lsu_acsphoton2018}. 

\subsection*{Fabrication}
As discussed in the main text, the designs were fabricated on an AIM Photonics 300 mm wafer multi-project wafer (MPW) run \cite{aimphotonics}. We waived the minimum width DRC (design-rule check) rule for the spatial mode multiplexer, 50-50 directional coupler, and wavelength demultiplexer, and waived the minimum separation rule for the spatial mode multiplexer and wavelength demultiplexer. However, the designs were all successfully resolved by the $193~\mathrm{nm}$ immersion lithography used in the AIM Photonics process, which can produce features as small as $40~\mathrm{nm}$ across \cite{mtotzeck_nphoton2007}.

\subsection*{Measurement}
We measured the transmission through the devices by a home-built grating-coupling setup. The input and output fibers were mounted on flexture stages with piezo nanopositioners (Thorlabs NanoMax) and were 10 degrees from normal to the chip plane. We used a bandpass filtered supercontinuum laser (Fianium WhiteLase SC-400-4) as light source and an optical spectrum analyzer (Agilent 86140B) to measure the transmitted power spectra. The fibers were automatically aligned to maximize the transmitted power before each measurement. The power spectra of individual devices are reproducible to $\pm$ 0.5 dB. We obtain the transmission spectra of the devices by normalizing against the two straight waveguides on each chip. The transmission of these two straight waveguides have a variation of up to $\pm 0.6~\mathrm{dB}$ on all three chips and we use the average spectra to normalize. In the figures containing experimental transmission data, the solid lines are the average spectra and the shaded regions the span of values from all three chips. 

\subsection*{Extracting S-parameters for spatial mode demultiplexer}
Extracting the S-parameters for a single multiplexer from the measured data is relatively straightforward. The $\mathrm{S}_{13}$ and $\mathrm{S}_{24}$ parameters of the back-to-back structure are a measure of insertion loss, and are equal to double the insertion loss of a single multiplexer. Meanwhile, the $\mathrm{S}_{14}$ and $\mathrm{S}_{32}$ parameters are measures of the crosstalk. In this test structure, there are two dominant crosstalk paths: light can be coupled into the wrong mode of the multi-mode waveguide by the first multiplexer, and light can be coupled into the wrong output waveguide by the second multiplexer. Since the $\mathrm{TE}_{00}$ and $\mathrm{TE}_{10}$ modes of the multi-mode waveguide have different propagation constants, this results in a fringing pattern in the $\mathrm{S}_{14}$ and $\mathrm{S}_{32}$ spectra. 

\section*{Acknowledgements}
This work was supported by AIM Photonics, and sponsored by the Air Force Research Laboratory via agreement number FA8650-15-2-5220, the Air Force Office of Scientific Research MURI on Attojoule Optoelectronics via award number FA9550-17-1-0002, and the Gordon and Betty Moore Foundation via grant GBMF4744.

\bibliography{invdes_references}

\end{document}